\documentstyle[12pt]{article}

\begin{document}

\begin{flushright}
IMSc/2009/05/08
\end{flushright} 

\vspace{2mm}

\vspace{2ex}

\begin{center}

{\large \bf Anisotropic Cosmology and (Super)Stiff Matter} \\ 

\vspace{2ex}

{\large \bf in Ho\v{r}ava's Gravity Theory }

\vspace{8ex}

{\large S. Kalyana Rama}

\vspace{3ex}

Institute of Mathematical Sciences, C. I. T. Campus, 

Tharamani, CHENNAI 600 113, India. 

\vspace{1ex}

email: krama@imsc.res.in \\ 

\end{center}

\vspace{6ex}

\centerline{ABSTRACT}
\begin{quote} 

We study anisotropic cosmology in Ho\v{r}ava's gravity theory
and obtain Kasner type solutions, valid for any number $d$ of
spatial dimensions. The corresponding exponents satisfy two
relations, one involving the marginal coupling $\lambda
\;$. Also, Ho\v{r}ava's (super)renormalisable theory predicts
(super)stiff matter whose equation of state is $p = w \rho \;$
with $w \ge 1 \;$. We discuss briefly the implications of these
results for the nature of cosmological collapse.

\end{quote}

\vspace{2ex}







\newpage

\vspace{2ex}

\centerline{ {\bf 1. Introduction} }

\vspace{2ex}

Recently, Ho\v{r}ava has proposed \cite{h1} a candidate theory
for gravity based on anisotropic scaling of space and time
coordinates: 
\begin{equation}\label{z}
x^i \; \to \; l x^i
\; \; \; , \; \; \; 
t \; \to \; l^z t 
\end{equation}
where $z$ is the scaling exponent. He has constructed an action
for the metric fields invariant under the above scaling and also
under foliation preserving diffeomorphic transformations. The
action is required to have no more than two time derivatives.
The kinetic part of the action is then universal and is
characterised by a marginal coupling $\lambda \;$. The potential
part has numerous terms containing various powers and spatial
derivatives of curvatures of the spatial metric. Ho\v{r}ava has
invoked `the principle of detailed balance' to constrain such
terms, but this seems unnecessary and may even be problematic.
The action reduces to the Einstein action in the infrared (IR)
if $\lambda \to 1 \;$ and $z \to 1 \;$.  The theory then has
full space--time diffeomorphism symmetry and may, therefore, be
a candidate for a renormalisable Einstein's theory of gravity.
Ho\v{r}ava's theory may also acquire an anisotropic Weyl
symmetry at $\lambda = \frac{1}{d} \;$ where $d$ is the number
of spatial dimensions.

Such a theory has many appealing properties. For example, it is
ghost free since there are no more than two time derivatives.
By construction, it is power counting renormalisable in the
ultraviolet (UV) if $z = d$ and is super renormalisable if $z >
d$, so it is believed to be UV complete. It contains many higher
powers and derivatives of curvature, hence it may be able to
resolve singularities. It singles out time, so the causal
structure in UV is likely to be modified which may have non
trivial implications for black hole physics. In such a theory,
the speed of light generically diverges in the UV, so the
horizon problem may perhaps be solved without requiring
inflation. See \cite{h1, h2} for more details.

Various aspects of such a theory are being studied actively.
See, for example, \cite{h2} -- \cite{nishioka}. In this letter,
we focus on implications of such a theory in early universe
cosmology where differences from Einstein's theory are likely to
be manifest. Such implications have also been studied in
\cite{gc, kiritsis, mukohyama, branden, piao, gao} for $d = 3$
homogeneous isotropic FRW universe. It is found that scale
invariant cosmological perturbations can be generated without
requiring inflation if the scale factor $a(t)$ evolves as $\sim
t^n \;$ with $n > \frac{1} {3} \;$; it is also found that there
can be a bounce in the early universe if the spatial curvature
is non zero. The scale invariance of perturbations is due to the
modifications of dispersion relations arising from anisotropic
scaling symmetry and, hence, is likely to be a generic feature
of Ho\v{r}ava's theory. The bounce is due to non zero spatial
curvature of the FRW universe and due to higher powers of
curvature in the action. The bounce is thus a possible but non
generic feature of Ho\v{r}ava's theory. For example, it is
absent for spatially flat universe.

Our main motivation here is to find the implications of
Ho\v{r}ava's theory which differ from those of Einstein's theory
and which are not crucially dependent on the spatial curvature.
We find that generic equation of state for the matter in the UV
is $\frac{p}{\rho} = w = \frac{z}{d} \;$. This is independent of
whether the spatial curvature is zero or non zero. Thus, in the
early universe, $w = 1$ for renormalisable theory whereas $w >
1$ for super renormalisable theory. The corresponding matter is
sometimes referred to as (super)stiff.

We consider the general action given in \cite{kiritsis} and
study the evolution of a homogeneous anisotropic universe with
$d$ spatial dimensions. We obtain anisotropic Kasner type
solution in the limit where universe is collapsing to zero
size. The exponents in the corresponding scale factors satisfy
two relations, one of them involving $\lambda \;$. The marginal
coupling $\lambda \;$ may be different from $1$ in the UV, and
may even be close to $\frac{1} {d} \;$ where the theory may
acquire a Weyl symmetry.  Such a behaviour of $\lambda \;$ and
the presence of (super)stiff matter have interesting
implications for the nature of collapse which we explain
briefly.

This paper is organised as follows. In section {\bf 2} we
present the set up. In section {\bf 3} we discuss the dispersion
relation and the consequent equation of state. In section {\bf
4} we present the equations of motion and some solutions, and
discuss briefly their implications. In section {\bf 5} we
conclude with a brief summary and a few comments.

\vspace{2ex}

\centerline{ {\bf 2. Ansatz for action and metric} }

\vspace{2ex}

In Ho\v{r}ava's theory, the fields are the lapse function $N$,
shift vector $N^i \;$, and the spatial metric $g_{i j} \;$. The
scaling dimensions of various quantities in momentum units are:
\[
[x^i] = - 1 \; \; , \; \; \; 
[t] = - z \; \; , \; \; \; 
[N] = [g_{i j}] = 0 \; \; , \; \; \; 
[N^i] = z - 1 \; \; . 
\]
The action $S = S_K + S_V$ is required to contain no more than
two time derivatives, and to be invariant under the scaling in
equation (\ref{z}) and foliation preserving diffeomorphism. The
kinetic part $S_K$ of the action is then universal and may be
written as 
\begin{equation}\label{sk}
S_K = \frac{1}{2 \kappa^2} \; \int \; d t \; d^d x \; N \sqrt{g}
\; \left( K_{i j} K^{i j} - \lambda K^2 \right)
\end{equation}
where $\kappa^2 \;$ is a parameter with dimension $[\kappa^2] =
z - d \;$, $\lambda \;$ is a dimensionless parameter, the
spatial indices $i, j, \cdots = 1, 2, \cdots, d \;$ are to be
lowered or raised using $g_{i j} \;$ or its inverse $g^{i j}
\;$,
\[
K_{i j} = \frac{1}{2 N} \; \left( \dot{g}_{i j} 
- \nabla_i N_j - \nabla_j N_i \right)
\; \; , \; \; \; 
K = g^{i j} \; K_{i j} \; \; ,
\]
and the covariant derivatives, as well as curvature tensors
below, are all with respect to $g_{i j} \;$. For $z = d \;$, the
parameter $\kappa \;$ becomes dimensionless and the theory is
power counting renormalisable; for $z > d \;$ it is super
renormalisable \cite{h1}. Our interest is in the $d = 3 \;$
case, but most of the expressions below are valid for any value
of $d \;$.

The potential part $S_V$ of the action contains various powers
and spatial derivatives of the Riemann tensor $R_{i j k l} \;$,
equivalently of the Ricci tensor $R_{i j} \;$ in the $d = 3 \;$
case. It suffices our purposes to write $S_V \;$ symbolically as
\begin{equation}\label{sv}
S_V = \int \; d t \; d^d x \; N \sqrt{g} \;
\left( \sigma + \xi \; R + \sum_{n = 2}^{n_*} \zeta_n \; R^n
+ \sum_{p, q = 1}^{p_*, q_*} \beta_{p q} R \nabla^p R^q \right)
\end{equation}
where the first sum denotes various powers of curvature tensor
and the second denotes various derivatives acting on various
powers of curvature tensor. The upper limits of $(n, p, q) \;$
depend on the value of $z \;$. For the renormalisable case, for
example, $z = d \;$ and $n_* = z \;$, $p_* + 2 q_* + 2 = 2 z
\;$. 

In \cite{h1}, Ho\v{r}ava invokes `the principle of detailed
balance' which will constrain the above form for $S_V \;$. For
example, in $d = 3 \;$ case, $S_V$ will not contain $R^3$ terms
and the coefficients of various terms in $S_V$ depend only on
three new parameters. By construction, the resulting action is
not the most general one. However, quantum corrections may not
obey the principle of detailed balance and may induce other
possible terms. This principle may even be problematic since the
corresponding static spherically symmetric solutions reduce to
the IR ones only on scales beyond the cosmological horizon, and
not on smaller scales where Einstein's theory has been well
tested. To rectify this problem requires going beyond the
detailed balance. See \cite{gc, kiritsis, pope} and, in
particular, \cite{nastase} for detailed discussions of these
issues.

For these reasons, we will not invoke detailed balance in this
paper, and consider the general form for the action $S_V$. The
most general form of $S_V$ for $z = d = 3$ is given, for
example, in \cite{kiritsis} where the corresponding equations of
motion are also obtained. These equations are very long and,
hence, are not presented here but will be used for the present
cosmological study. Note that, for such a study, one can set $N
= 1 \;$ and $N^i = 0 \;$ in the equations of motion with no loss
of generality; and, also that these equations are applicable for
any value of $d \;$ as we later explain.

Here, we consider only spatially curved, homogeneous, isotropic
universe; or spatially flat, homogeneous, anisotropic universe.
The line element of a spatially curved, homogeneous, isotropic
universe may be written as
\begin{equation}\label{dsiso}
d s^2 = - d t^2 + a^2 \; d \Sigma_{d, \hat{k}}^2
\end{equation}
where $a(t) \;$ is the scale factor, $d \Sigma_{d, \hat{k}} \;$
is the line element of a $d$ -- dimensional space of constant
curvature, and $\hat{k} = + 1, \; - 1, \; 0 \;$ for positive,
negative, or zero curvature. The Hubble parameter $H$ is defined
by $H = \frac{\dot{a}} {a} \;$ where an overdot denotes time
derivative.

The line element of a spatially flat, homogeneous, anisotropic
universe may be written as
\begin{equation}\label{dsaniso}
d s^2 = - d t^2 + \sum_{i = 1}^d a_i^2 \; (d x^i)^2
\end{equation}
where $a_i(t) \;$ are the scale factors.  The corresponding
Hubble parameters $h_i$ are defined by $h_i = \frac{\dot{a}_i}
{a_i} \;$. Also, define the geometric mean $a$ of the scale
factors by $a^d = \prod_i a_i \;$. Then $H = \frac{\dot{a}} {a}
= \frac{1}{d} \; \sum_i h_i \;$ is the average of the Hubble
parameters $h_i \;$.

With the above definitions, the conservation equation for a
matter source with pressure $p$ and density $\rho \;$ is given
in both of the above cases by
\begin{equation}\label{rhot} 
\dot{\rho} + d \; H \; (\rho + p) = 0 \; . 
\end{equation}
If the equation of state is given by $p = w \; \rho \;$ then we
have $\rho = \rho_0 \; a^{- d (1 + w)} \;$ where $\rho_0$ is an
initial value.

\vspace{2ex}

\centerline{ {\bf 3. (Super)stiff matter } }

\vspace{2ex}

Consider now matter sources, {\em e.g.}  radiation, and their
equations of state. The matter action which is invariant under
the scaling in equation (\ref{z}) will lead to a modified
dispersion relation in the UV, typically of the form $\omega^2
\sim k^{2 z} \;$ \cite{h1, h2}. From the standard statistical
mechanical methods using such a dispersion relation, it follows
that the dependence of free energy $F$ on temperature $T$ is of
the form $F \sim T^{1 + \frac{d}{z}} \;$ \cite{h2, chen}. For
renormalisable theories $z = 1 \;$ in the IR and $z = d \;$ in
the UV. It then follows that $F \sim T^{1 + d} \;$ at low
temperatures and $F \sim T^{1 + 1} \;$ at high temperatures. As
noted in \cite{h2}, similar free energy behaviour at high
temperature appears also in string theory. We further note here
that similar free energy behaviour, at both low and high
temperatures, appears also in the context of a particular
version of generalised uncertainty principle \cite{gup}.

With free energy $F \sim T^{1 + \frac{d}{z}} \;$, it follows
upon using thermodynamical relations that the corresponding
equation of state is given by $p = w \; \rho$ where $w =
\frac{z}{d} \;$. Thus for radiation in $d = 3 \;$, we have $z =
1 \;$ and $w = \frac{1}{3} \;$ in IR. We have $z = d \;$ in the
UV for renormalisable theories which then implies that $w = 1
\;$, \footnote{That $w = 1 \;$ for radiation in the UV is also
pointed out in \cite{mukohyama2} which appeared while writing up
this paper.} the corresponding matter sometimes referred to as
stiff matter. Also, $z > d \;$ for super renormalisable theories
which then implies that $w$ can be $> 1 \;$, the corresponding
matter sometimes referred to as superstiff matter.

Such an UV dispersion relation, namely $\omega^2 \sim k^{2 z}
\;$, is ubiquitous in Ho\v{r}ava's theory and arises from an
underlying principle: it is a consequence of invariance under
the anisotropic scaling in equation (\ref{z}). Also, it is
independent of whether spatial curvature is zero or non
zero. Thus, Ho\v{r}ava's theory can be taken to predict that
early universe, and more generally UV regime, is dominated by
matter whose equation of state is given by $p = w \rho$ where $w
= \frac{z}{d} = 1$ for renormalisable theories and $> 1 \;$ for
super renormalisable theories. \footnote{ The idea that early
universe must be dominated by $w \ge 1$ matter also appears in
different contexts. For example, see \cite{bfw1, k3+1} for the
$w = 1$ case; see \cite{wesley} and references therein for the
$w > 1$ case.}

\vspace{2ex}

\centerline{ {\bf 4. Equations of motion, solutions and their
implications} }

\vspace{2ex}

Consider now equations of motion. They are given in
\cite{kiritsis} for $d = 3 \;$. Consider, for any $d \;$, the
general form of the contributions of various terms in the action
to the equations of motion. For the cases of interest here,
namely where the line element is given by equation (\ref{dsiso})
or (\ref{dsaniso}), we observe the following: \\

\noindent
{\bf (i)} 
Matter source with an equation of state $p = w \; \rho \;$ will
contribute terms $\propto a^{- d (1 + w)} \;$ in the equations
of motion, see equation (\ref{rhot}). \\

\noindent 
{\bf (ii)} 
Consider terms of the form $R^n \;$ in $S_V \;$. For the
isotropic case, it is easy to see that they contribute a term
$\propto \hat{k}^n \; a^{- 2 n} \;$ in the equations of
motion. It thus follows that such terms act as sources with
equations of state $p = w \; \rho \;$ where $w = \frac{2 n}{d} -
1 \;$ and $\rho = C_n \hat{k}^n \; a^{- 2 n} \;$. The constant
$C_n$ depends on the index structure of $R^n \;$ terms and their
coefficients in $S_V \;$. For the spatially flat case, $\hat{k}
= 0 \;$ and the corresponding contributions all vanish.

Note that $n = 1$ for $R$ term and $w = \frac{2}{d} - 1 = -
\frac{1} {3} \;$ for $d = 3$; $n = 2$ for $R^2 \;$ term and $w =
\frac{4}{d} - 1 = \frac{1}{3} \;$ for $d = 3$; and, formally, $n
= 0$ for cosmological constant term and $w = - 1 \;$ for any
value of $d \;$. For the term $R^{n_*} \;$ with highest power of
curvature, we have $n_* = z = d \;$ for the renormalisable case
and $w = \frac{2 n_*}{d} - 1 = 1 \;$ for any value of $d \;$. \\

\noindent
{\bf (iii)} 
$R_{i j k l} \;$ for a constant curvature space is given in
terms of $g_{i j} \;$, and the scale factor which depends on $t$
only. Hence covariant derivatives acting on curvature tensors
will all vanish. Therefore the terms in the second sum in
equation (\ref{sv}) do not contribute to the equations of
motion. \\

\noindent
{\bf (iv)}
The kinetic part $S_K \;$of the action is universal for any
values of $d \;$ and $z \;$. Hence, the corresponding terms in
the equations of motion are just those given in 
\cite{kiritsis}. \\

Using the observations {\bf (i) -- (iv)} above and the
expressions given in \cite{kiritsis}, we can now write the
equations of motion.  

\vspace{2ex}

\centerline{ {\bf Isotropic case} }

\vspace{2ex}

For the isotropic case, the metric is given in equation
(\ref{dsiso}) and the equations of motion may be written as
\begin{eqnarray}
d \; (\lambda d - 1) \; H^2 & = & 2 \kappa^2 \; \sum \rho
\label{iso1} \\
(\lambda d - 1) \; \left( \dot{H} + d H^2 \right) & = & \kappa^2
\; \sum (\rho - p) \label{iso2}
\end{eqnarray}
where $H = \frac{\dot{a}}{a} \;$ and $\dot{H} = \frac{\ddot{a}}
{a} - \left(\frac{\dot{a}}{a}\right)^2 \;$. The sum $\sum$ in
the equations above denotes contributions from matter source,
and also those from $R^n \;$ terms in $S_V \;$ for which $p =
\left( \frac{2 n}{d} - 1 \right) \;\rho \;$ and $\rho = C_n \;
\hat{k}^n \; a^{- 2 n} \;$ where the constant $C_n$ depends on
the index structure of $R^n \;$ terms and their coefficients in
$S_V \;$. See \cite{kiritsis} for explicit expressions for the
$d = 3$ case.

We will assume in this paper that $\lambda d > 1 \;$, since this
is the case in Einstein's theory for which $\lambda = 1 \;$, and
also that $\lambda$ is not arbitrarily close to $\frac{1}{d}
\;$. Replacing $\kappa^2 \;$ by $\left( \frac{\lambda d - 1} {d
- 1} \right) \kappa^2 \;$ then renders equations (\ref{iso1})
and (\ref{iso2}) identical to those in Einstein's theory.

The evolution of the scale factor is then straightforward to
understand. Let $a \to 0 \; \; (\infty) \;$ in the limit $t \to
0 \; \; (\infty) \;$. The evolution is then dictated by those
sources for which $w \;$ is largest (smallest). If the total
coefficient of the dominant sources is positive then it follows
that $a(t) \sim t^{\frac{2}{d (1 + w)}} \;$.

If the spatial curvature is non zero then sources arising from
curvature terms can have negative coefficients. Then the total
coefficient of the dominant sources can be negative and this
will generically lead to a bounce in the evolution of $a(t)
\;$. The details, and even the presence itself, of the bounce
depend on the nature and strength of other sources present and
can only be obtained by further analysis incorporating these
data.

Let $z = d = 3 \;$. Ho\v{r}ava's theory then predicts the
existence of stiff matter in the UV for which $w = 1 \;$.
Assume the total coefficient of the sources with $w = 1 \;$,
which are the dominant ones in the limit $a \to 0 \;$, to be
positive. This will be the case for spatially flat universe for
which there are no contributions from curvature terms. It then
follows that there is no bounce and that $a(t) \sim
t^{\frac{1}{3}} \;$ in the limit $t \to 0 \;$.

We now make a remark. It has been shown in \cite{kiritsis,
mukohyama, piao, gao} that, in Ho\v{r}ava's theory, scale
invariant primordial perturbation spectrum can be generated in
the UV with an additional scalar field and {\em without
requiring inflation}.  Scale invariance arises, essentially,
from the dispersion relation for the scalar field in the UV
which is of the form $\omega^2 \sim k^6 \;$. For the desired
dynamics of the perturbations thus generated, it is also
required that $H^2 a^6 \;$ be an increasing function of $t \;$,
or equivalently that $\int^\infty \frac{d t}{a^3} \;$ converge,
which is taken to imply that the scale factor $a \;$ evolve as
$\sim t^n \;$ with $n > \frac{1}{3} \;$. See \cite{kiritsis,
mukohyama, piao, gao} for details.

However, as described above, it is likely that $a(t) \sim
t^{\frac{1}{3}} \;$ in the UV. Although this violates the
requirement $n > \frac{1}{3} \;$, there may be no adverse effect
on scale invariance of the spectrum since $H^2 a^6 \;$ may still
be an increasing function of $t \;$ because of the presence of
other sources in equation (\ref{iso1}) which will become
important as $t$ increases. This is plausible but, nevertheless,
it is desireable to study in detail the effects of $a(t) \sim
t^{\frac{1}{3}} \;$ in the UV on the scale invariance of the
spectrum obtained in \cite{kiritsis, mukohyama, piao, gao} in
Ho\v{r}ava's theory without requiring inflation.

\vspace{2ex}

\centerline{ {\bf Anisotropic case} }

\vspace{2ex}

For the spatially flat anisotropic case, the metric is given in
equation (\ref{dsaniso}). There are no contributions from $R^n
\;$ terms and the equations of motion may be written as
\begin{eqnarray} 
\lambda \; d^2 H^2 - \sum_i (h_i)^2 & = & 2 \kappa^2 \; \rho
\label{aniso1} \\ 
(\lambda d - 1) \; \left( \dot{h}_i + d H h_i \right) & = &
\kappa^2 \; (\rho - p) \label{aniso2} 
\end{eqnarray}
where $h_i = \frac{\dot{a}_i}{a_i} \;$, $H = \frac{1}{d} \;
\sum_i h_i \;$, and $\dot{h}_i = \frac{\ddot{a}_i}{a_i} - \left(
\frac{\dot{a}_i}{a_i} \right)^2 \;$. Note that summing equation
(\ref{aniso2}) over $i \;$ gives
\begin{equation}\label{aniso3} 
(\lambda d - 1) \; \left( \dot{H} + d H^2 \right) = \kappa^2 \;
(\rho - p) \; \; .
\end{equation}
We have $H = \frac{\dot{a}}{a} \;$ from the definition $a^d =
\prod_i a_i \;$. It then follows that \footnote{Equations
(\ref{aniso2}) and (\ref{aniso3}) give equation (\ref{lit}).
Using $\sum_i h_i = d H \;$ gives the constraint $\sum_i A_i = 0
\;$. Substituting $h_i \;$ in equation (\ref{aniso1}) then gives
equation (\ref{Lt2}).} 
\begin{eqnarray} 
h_i - H & = & A_i \; a^{- d} \label{lit} \\ 
d \; (\lambda d - 1) H^2 & = & 2 \kappa^2 \; \rho + A^2 \; 
a^{- 2 d} \label{Lt2}
\end{eqnarray} 
where $A_i \;$ are initial values satisfying $\sum_i A_i = 0 \;$
and $A^2 = \sum_i (A_i)^2 \;$. Once the equation of state
$p(\rho) \;$ is given then, in principle, $\rho(a) \;$ can be
obtained from equation (\ref{rhot}), $a(t) \;$ from equation
(\ref{Lt2}), and $a_i(t) \;$ from equation (\ref{lit}). Note
that the initial values $A_i$ encode anisotropic initial
conditions, {\em e.g.} during a collapse, and also that $A_i$
may be thought of as a source with equation of state $p = w \;
\rho \;$ where $w = 1$ and $\rho_0 = \frac{A^2}{2 \kappa^2} \;$,
see equations (\ref{rhot}) and (\ref{Lt2}).

Consider now the dynamics of the evolution, assuming the
equation of state to be $p = w \rho \;$. Equation (\ref{rhot})
then implies that $\rho = \rho_0 \; a^{- d (1 + w)} \;$ where
$\rho_0 > 0 \;$ is an initial value. The evolution in the limit
$a \to \infty \;$, namely large universe limit, is similar to
the standard one where $a \sim t^{\frac{2}{d (1 + w)}} \;$. The
effect of $\lambda \;$ is unimportant unless $\lambda$ is
arbitrarily close to $\frac{1}{d} \;$.

Consider a universe collapsing to zero size, {\em i.e.} $a \to 0
\;$, as $t \to 0 \;$. Let the scale factors $a_i \sim
t^{\alpha^i}$ in this limit. We study the following cases.

\vspace{2ex}

\centerline{{\bf $w > 1$}} 

\vspace{2ex}

In the limit $a \to 0 \;$, $2 \kappa^2 \rho \sim a^{- d (1 + w)}
\; \gg A^2 a^{- 2 d} \;$ in equation (\ref{Lt2}) since $w > 1
\;$. It is then straightforward to show that 
\begin{equation}\label{w>1}
a \sim t^{\frac{2}{d (1 + w)}} \; \; \; , \; \; \;
a_i = c_i \; e^{c \; \; t^{\frac{w - 1}{w + 1}}} \; 
t^{\frac{2} {d (1 + w)}}
\end{equation}
where $c_i$ and $c$ are constants. Thus, since $t^{\frac{w - 1}
{w + 1}} \to 0 \;$ in the limit $t \to 0 \;$, it follows that
the exponents $\alpha^i \;$ in $a_i \sim t^{\alpha^i} \;$ are
all equal, and are independent of the initial values $A^i
\;$. Hence, the collapse is isotropic and stable under
perturbations \cite{wesley}. 

\vspace{2ex}

\centerline{{\bf $w \le 1$}} 

\vspace{2ex}

In the limit $a \to 0 \;$, the right hand side of equation
(\ref{Lt2}) becomes $B^2 a^{- 2 d} \;$ where $B^2 = A^2 \;$ if
$w < 1$ and $B^2 = 2 \kappa^2 \rho_0 + A^2 \;$ if $w = 1 \;$. It
is then straightforward to show that
\begin{equation}\label{wle1}
a \sim t^{\frac{1}{d}} \; \; , \; \; \;
a_i \sim t^{\alpha^i} \; \; , \; \; \; 
\alpha^i = \frac{1}{d} - \frac{A^i}{B} \; 
\left( \lambda - \frac{1} {d} \right)^{\frac{1}{2}} \; \; .
\end{equation}
This is a Kasner type solution. The exponents $\alpha^i$ depend
on initial values $A^i \;$ and, since $\sum_i A^i = 0 \;$,
satisfy the relations
\begin{equation}\label{kasner}
\sum_i \alpha^i = 1
\; \; , \; \; \; 
X \equiv \sum_i (\alpha^i)^2 = \frac{1}{d} + \left( \lambda 
- \frac{1}{d} \right) \; \frac{A^2}{B^2} \; \; .
\end{equation}
For $w < 1 \;$ we have $B^2 = A^2 \;$, hence $X = \lambda \;$ is
the only possible value. For $w = 1 \;$ we have $B^2 = 2
\kappa^2 \rho_0 + A^2 \;$, hence $0 < \frac{A^2}{B^2} < 1 \;$
and $\frac{1}{d} < X < \lambda \;$. Clearly $X \to \lambda \;$
if $2 \kappa^2 \rho_0 \ll A^2 \;$. Also $X \to \frac{1}{d} \;$
if $2 \kappa^2 \rho_0 \gg A^2 \;$ which is the only possibility
in Einstein's theory, or if $\lambda \to \frac{1}{d} \;$ which
is a new possibility in Ho\v{r}ava's theory and is valid for any
values of $\rho_0$ and $A^2 \;$.

Kasner type solutions, in particular the value of $X \;$ and the
dependence of $\alpha^i \;$ on initial values, provide an
insight into the stability of the cosmological collapse process
under generic curvature and/or anisotropic perturbations;
namely, an insight into whether the collapse will be isotropic
or anisotropic, whether smooth or will exhibit chaotic
oscillatory behaviour, et cetera.

Since $\alpha_i$ depend on the initial values $A_i \;$, the
collapse will be generically anisotropic. Consider $d = 3 \;$
case. The exponents $\alpha_i \;$ must satisfy the constraints
in equation (\ref{kasner}). If $X = 1 \;$ then one of the
$\alpha_i$ must be negative. Then, under curvature
perturbations, the collapse will not be smooth and will exhibit
chaotic behaviour.  If $X \;$ is sufficiently close to
$\frac{1}{3} \;$ then no $\alpha_i \;$ can be negative and the
collapse will be stable and non oscillatory under perturbations.

In Einstein's theory $\lambda = 1$ and, hence, smaller values of
$X \;$ may result only through smaller values of $\frac{A^2}{2
\kappa^2 \rho_0} \;$ which necessarily requires stiff matter
with $w = 1 \;$. Stability also results if superstiff matter
with $w > 1 \;$ is present as follows from equation (\ref{w>1}).
See \cite{wesley} and the references therein for a thorough
discussion of these issues. 

In Ho\v{r}ava's theory, on the other hand, $\lambda \;$ is
typically different from $1 \;$ in the UV. This theory may
acquire an anisotropic Weyl symmetry if $\lambda = \frac{1}{d}
\;$, so it is possible that $\lambda \to \frac{1}{d} \;$ in the
UV. If so then $X$ may be naturally small even without stiff
matter. But (super)stiff matter with $w \ge 1$ is also naturally
present in this theory. This makes it more likely that collapse
process is stable and non oscillatory.

However, spatial curvature terms of high order are also allowed
in Ho\v{r}ava's theory. As explained in remark {\bf (ii)} in
section {\bf 4}, this order is closely linked to the scaling
exponent $z$, to which is also linked the presence of
(super)stiff matter. Curvature terms typically lead to
destabilising effects under curvature perturbations and our
preliminary analysis indicates that they may be comparable to
the stabilising effects of (super)stiff matter. However, there
is also the stabilising effect that results if $\lambda \;$ is
close to $\frac{1}{d} \;$ in the UV but no comparable,
$\lambda-$dependent, destabilising curvature effects seem to be
present. It is therefore possible that the sum total of all
these effects in Ho\v{r}ava's theory results in a stable and non
oscillatory collapse. Clearly, further analysis is necessary but
it is complicated and quite involved, and is beyond the scope of
the present work.

\vspace{2ex}

\centerline{ {\bf 5. Conclusion} }

\vspace{2ex}

We now summarise briefly and make a few comments. Our main
motivation for the present study is to find the implications of
Ho\v{r}ava's theory which differ from those of Einstein's theory
and which are not crucially dependent on the spatial curvature.
The implications of Ho\v{r}ava's theory we find which differ
from those of Einstein's theory are: {\bf (1)} The UV regime in
the (super)renormalisable case is dominated by (super)stiff
matter, namely matter with $w \ge 1 \;$. {\bf (2)} $\sum_i
(\alpha^i)^2 \;$, where $\alpha^i$ are the Kasner exponents, can
be different from and smaller than $1$ even without stiff matter
present. These two implications are generic. {\bf (3)} Equations
of motion contain curvature terms of the form $C_n \; \hat{k}^n
\; a^{- 2 n} \;$ where $n \le d$ for the renormalisable case.
The constants $C_n \;$ can be positive or negative. The
curvature terms may hence lead to a bounce in the evolution of
the scale factor $a(t) \;$. The bounce is possible but non
generic since it depends on the sign and the magnitude of $C_n
\; \hat{k}^n \;$ as well as the nature and strength of other
sources present.

We now make a few comments. Ho\v{r}ava's theory is believed to
be UV complete. Also, it contains higher powers and derivatives
of curvature. It is then reasonable to expect that this theory
will also resolve the singularities. The big bang singularity is
indeed absent if $a(t) \;$ bounces back. But the bounce is not a
generic feature. If there is no bounce then, as can be inferred
from the present solutions, the big bang singularity is
present. Note that black hole singularity is also present in all
the static spherically symmetric solutions to Ho\v{r}ava's
theory studied so far, see {\em e.g.} \cite{pope, nastase, cai,
cai2, colgain}, although these solutions do differ from those in
Einstein's theory.

It is possible that, regarding the presence or absence
singularities, this is the most one can see using classical
action in Ho\v{r}ava's theory and that quantisation of the
action is necessary to see any further.

There is a similar situation in string/M theory. No classical
string/M theory action so far has led to generic absence of
singularities. However, it is likely that when the temperatures
become comparable to string theory scales the classical
description of the universe, and similarly of black holes, must
be abandoned and string theoretic description should be used,
see \cite{bowick, hp}. But the details of such a description, or
of how exactly the singularities get resolved, are not fully
known at present. However, due to entropic reasons, the universe
at this stage seems likely to be dominated by stiff matter for
which $w = 1 \;$ \cite{bfw1, k3+1}. \footnote{The nature and
quantum mechanical properties of stiff matter, and those of a
universe dominated by stiff matter, and further evolution of
such a universe are studied in a series of papers in
\cite{bfw1}. A possible string/M theory scenario of how such a
universe may arise is also given in \cite{k3+1}. Also, the
properties of stars made up of stiff matter and their
similarities to black holes are studied in \cite{bfstiff}.}

In this context, note that Ho\v{r}ava's theory also predicts
generically the presence of stiff matter in the early universe.
The presence of such matter and/or the scaling arguments using
equation (\ref{z}) lead to the high temperature behaviour of the
free energy $F \sim T^2 \;$ \cite{h2}. This behaviour may be
taken to signify that the spectral dimension of the spacetime in
the UV is $1 + 1 \;$. This is shown to be the case for
Ho\v{r}ava's theory in \cite{h2}. Such an UV spectral dimension
is also observed in many candidate theories for quantum gravity
{\em e.g.} causal dynamical triangulations, quantum Einstein
gravity, spin foam theory, and string theory. This may perhaps
be the case also for loop quantum gravity. It is indeed argued
in \cite{gc} that two dimensional effective gravitational
theories in the UV may be a generic feature of UV complete
theories of gravity. See \cite{gc} for more discussions and an
extensive list of references.

If these similarities are more than just coincidences then it
may be that there are also similarities in the ways in which
singularities get resolved in any of these theories and in
Ho\v{r}ava's theory. It is therefore important to study if
singularities can be resolved by quantising the action in
Ho\v{r}ava's theory. Such a resolution, besides being important
on its own, may also provide insights into the theories
mentioned above.


\vspace{3ex}

{\bf Acknowledgement: } We thank the referee for many helpful
suggestions which improved the presentation of the paper.



\begin{thebibliography}{999}
\bibitem{gup}
Modified dispersion relations appear in other contexts also, but
depend on an arbitrary choice of certain function(s). In the
context of a particular version of generalised uncertainty
principle, such a choice is fixed by a set of group theoretic
assumptions \cite{maggiore}. The corresponding statistical
mechanics incorporating the generalised uncertainty principle is
then shown in \cite{kgup} to lead to a free energy with the same
behaviour as appears here for renormalisable theories -- namely,
$F \sim T^{1 + d} \;$ at low temperatures and $F \sim T^{1 + 1}
\;$ at high temperatures.
\end{thebibliography}
\end{document}